\documentclass[english]{article}
\usepackage[T1]{fontenc}
\usepackage[utf8]{luainputenc}
\usepackage{geometry}
\usepackage{array}
\usepackage{float}
\usepackage{multirow}
\usepackage{amsmath}
\usepackage{amssymb}
\usepackage{graphicx}

\makeatletter

\providecommand{\tabularnewline}{\\}

\newenvironment{lyxlist}[1]
{\begin{list}{}
{\settowidth{\labelwidth}{#1}
 \setlength{\leftmargin}{\labelwidth}
 \addtolength{\leftmargin}{\labelsep}
 }}
{\end{list}}

\makeatother

\usepackage{babel}
\begin{document}

\title{Automatic parametrization of age/ sex Leslie matrices for human populations
(draft)}

\author{Webb Sprague, webbs@demog.berkeley.edu%
\thanks{I would like to thank the following for invaluable assistance: Professors
Jenna Johnson-Hanks, Ronald Lee, and Ken Wachter (UC Berkeley), Dr.
Jack Backer (UNM), and Eddie Hunsinger (Alaska state demographer).
I have received many helpful comments from fellow students in the
Demography 296 Seminar. I would also like to thank Professor Loo Botsford
(UC Davis), who first introduced me to Leslie matrices and mentioned
the potential of optimization in population biology in 2001.%
}}
\maketitle
\begin{abstract}
In this paper, we present a technique for generating Leslie transition
matrices from simple age and sex population counts, using an implementation
of \textquotedbl{}Wood's Method\textquotedbl{} \cite{wood}; these
matrices can forecast population by age and sex (the \textquotedbl{}cohort
component\textquotedbl{} method) using simple matrix multiplication
and a starting population. Our approach improves on previous methods
for creating Leslie matrices in two respects: it eliminates the need
to calculate input demographic rates from \textquotedbl{}raw\textquotedbl{}
data, and our new format for the Leslie matrix more elegantly reveals
the population's demographic components of change (fertility, mortality,
and migration). The paper is organized around three main themes. First,
we describe the underlying algorithm, \textquotedbl{}Wood's Method,\textquotedbl{}
which uses quadratic optimization to fit a transition matrix to age
and sex population counts. Second, we use demographic theory to create
constraint sets that make the algorithm useable for human populations.
Finally, we use the method to forecast 3,120 US counties and show
that it holds promise for automating cohort-component forecasts. This
paper describes the first published successful application of Wood's
method to human populations; it also points to more general promise
of constrained optimization techniques in demographic modeling.
\end{abstract}

\section{Introduction}

\subsection{Leslie matrices and cohort component forecasting}

\begin{flushleft}
Computing age and sex specific forecasts (ASSFs) is a standard task
of applied demographers, as these forecasts form a fundamental input
for planning throughout industry and government. Examples include
regular forecasts by the the U.S. Census Bureau, K-12 school forecasting,
and caseload projection (hospitals, prisons, medicaid, etc). As one
example, Washington State produces ASSFs as part of their statewide
land use planning program; these forecasts are used to plan for innumerable
downstream projects, from hospital construction to future transportation
projects. 
\par\end{flushleft}

\begin{flushleft}
The cohort component method invented by Whelpton \cite{whelpton28}
is a well established procedure for creating ASSFs. The Leslie matrix
formulation of the cohort component method is particularly elegant
for many reasons: it allows a forecasting problem to be formulated
as simple matrix multiplication, the matrix provides a succinct characterization
of the population dynamics, and the linear algebraic structure of
the matrix yields interesting results. However, creating Leslie matrices
requires demographic rates for the population under question, which
must either be drawn from model demographic rates or created directly
from vital event counts and age/sex-specific base population estimates.
Even when these prerequisites are met (which can be surprisingly difficult
in an applied demographic institutional environment), the Leslie matrix
generated usually requires extensive tuning to yield plausible forecasts,
because the results are highly sensitive to small changes in the matrix,
whether due to error or mis-estimation. Using Wood's method, however,
the Leslie matrix is guaranteed to generate at least plausible forecasts
insofar as it is forced to interpolate between the input population
vectors; in our experiments, it extrapolates populations well too.
In this paper we describe a procedure such that an applied demographer
can simply input historic age/sex population counts -- no need for
separate rate data -- and automatically get a plausible Leslie transition
matrix that can be used for cohort-component projection. 
\par\end{flushleft}

\begin{flushleft}
To our knowledge, the underlying algorithm in our technique -- Wood's
method (WM) -- has not been successfully applied to human demographic
problems, even though it has much promise for applied work. The lack
of uptake may be due to several reasons, not least of which is that
formulating the quadratic optimization problem in WM requires intermediate
skills in computer programming and access to a good matrix programming
language.%
\footnote{We would not consider SAS or a spreadsheet program to fulfill this
requirement, and these are the only environments one can count on
seeing in applied demography shops. R is an improvement, but R's syntax
is awkward for implementing complex matrix algorithms. The author
prefers Octave, a Matlab compatible language; if Octave is not available,
Matlab provides a reasonable second-best alternative. %
} More importantly, though, without well chosen constraints for the
optimization model, the method gives completely implausible results,
which might lead a researcher to abandon the technique prematurely.
In the only work on WM besides the original paper, Caswell's influential
book \cite{mpm}, Caswell recommends rather simplistically chosen
constraints that force the cell and row sums in the output Leslie
matrix to fall between zero and one; these constraints may be reasonable
for a closed single sex population, but when used to derive a Leslie
matrix for human populations with open migration flows, they yield
a matrix with no demographic interpretation and which gives very poor
forecasts. However, if constraints are developed using basic human
demographic theory, WM provides interpretable projection matrices
which yield very reasonable forecasts. 
\par\end{flushleft}

\begin{flushleft}
We will not do extensive comparisons to other forecasts in this paper,
either in terms of results or methodologies, but we plan to cover
both in \cite{sprague-projections}, where we will compare WM generated
forecasts to many official forecasts. As part of that paper, we will
also present results of an informal survey of applied demographers
regarding their forecasting methods and data sources. To our very
limited knowledge, most applied demographers depend either on spreadsheets
or SAS arrays to perform cohort component analysis and don't use matrix
methods at all. We hope that this paper will make these methods more
attractive by showing their power and elegance.
\par\end{flushleft}

\section{Method}

\begin{flushleft}
There are a number of moving parts to the method and we will take
each in turn: to begin, we will give a short introduction to the idea
of optimization under constraint, then we will describe the input
format to Wood's method, followed by the transition matrix format,
then the Woods inference algorithm, and finally a description of the
constraints. It will be helpful to have a basic overview of the method
before diving into the details: First, at least three sets of age/
sex population counts are assembled, with age widths and time periods
at 5 year intervals, and with 85 years old as the open interval.%
\footnote{This open interval chosen because age specific demographic data in
the United States is most commonly found with the open interval at
85 years old. This method is applicable to any regular age division;
in the future we will use it with population count data using 100+
as the open interval.%
} Second, a set of constraints is developed that sets bounds on the
survival, migration, and fertility cells in the output matrix, and
enforces any relationships between age-specific rates. Third, this
population count data and this constraint set are input to an Octave
function, which rearranges the population count data into a format
suitable for quadratic optimization. Fourth, this function calls the
function ``\texttt{qp()}'', which performs constrained quadratic
optimization on a decision vector in which each optimized element
is a structural non-zero cell in the desired Leslie matrix. Finally,
the resulting vector is rearranged into a Leslie matrix and returned.
Each of these steps will be described in detail below, but first we
will try to give some backround on optimization methods.
\par\end{flushleft}

\subsection{Optimization background}

\begin{flushleft}
As our presentation of WM highlights optimization techniques, it is
worth sketching those for the reader; please note, however, that it
would be impossible to do justice to this very well-developed topic
-- for an excellent introduction see \cite{optim}, but for a well
written and classic work see \cite{dantz}. Optimization models try
to find the best values for a set of variables under a set of constraints
as measured by a summarizing ``objective function'' -- for example,
finding the best diet as measured by total nutrition and constrained
by a budget, or the best allocation of retail products in a store
as measured by total profit and constrained by shelf space, or the
best fit of a parameterized function as measured by the minimum sum
of squares and constrained by bounds on the parameters (the optimization
problem used in WM). When the objective function is ``convex'' --
meaning roughly that no line that connects two points on the function
surface crosses the surface -- every local optimized value is guaranteed
to also be globally optimized value over the (usually constrained)
feasible solution space; linear functions are convex, and many second
order polynomials are convex as well, including the function corresponding
to sum of squares. 
\par\end{flushleft}

\begin{flushleft}
Most applied optimization problems are formulated in terms of strictly
linear equations, including the objective function, which is usually
a large summation involving the parameters and variables; this formulation
gives us the ``linear programming'' that underpins modern operations
research. In ``quadratic programming,'' the objective function is
a second order polynomial; quadratic programming is used in many variance
minimization problems including Markowitz portfolio optimization.
Optimization problems are ubiquitous in many fields, especially economics,
operations research, automatic controls, and machine learning -- even
modern cognitive linguistics -- and the algorithms and problem formulations
are very well developed. See below for a longer discussion of the
potential power of constrained optimization approaches in demographic
modeling beyond forecasting.
\par\end{flushleft}

\subsection{Population input format}

\begin{flushleft}
The input is a set of age sex population counts stored in a 36 by
T matrix, with the rows representing age and sex combinations and
the columns representing each five year interval. Male and female
population counts for the same year are concatenated vertically in
a single vector, and these vectors are concatenated horizontally into
a matrix; for an example see \eqref{eq:popvector}. At least three
years of data are required, but as many years of data as are available
can be used by the method. In \eqref{eq:popvector}, column 1 holds
the population for 1990, column 2 for 1995, column 3 for 2000.
\par\end{flushleft}

\begin{equation}
\begin{bmatrix}_{5}m_{0}^{1990} & _{5}m_{0}^{1995} & _{5}m_{0}^{2000}\\
_{5}m_{5}^{1990} & _{5}m_{5}^{1995} & _{5}m_{5}^{2000}\\
\vdots & \vdots & \vdots\\
_{+}m_{85}^{1990} & _{+}m_{85}^{1995} & _{+}m_{85}^{2000}\\
_{5}f_{0}^{1990} & _{5}f_{0}^{1995} & _{5}f_{0}^{2000}\\
_{5}f_{5}^{1990} & _{5}f_{5}^{1995} & _{5}f_{5}^{2000}\\
\vdots & \vdots & \vdots\\
_{+}f_{85}^{1990} & _{+}f_{85}^{1995} & _{+}f_{85}^{2000}
\end{bmatrix}\label{eq:popvector}
\end{equation}

\subsection{Output Leslie matrix }

\subsubsection{Single sex Leslie matrix review}

\begin{flushleft}
The Leslie matrix method should be familiar, at least in outline,
to analysts with formal demographic training; if not, excellent introductory
works include \cite{wachter-textbook}. However, it is worth reviewing
the essentials of the basic single sex Leslie matrix for a closed
population. Leslie matrices project a population vector like \eqref{eq:popvector-1}
forward by a single time step of the same length as the age classification
(e.g., if the population counts are reckoned in five year intervals,
then each projection step is five years long). The matrix multiplies
the population vector for time step 0 to yield a population vector
at time step 1: $\mathbf{p}_{1}=\mathbf{A}\mathbf{p}_{0}$, where
$\mathbf{p}_{1}$ is the projected age-specific population vector,
$\mathbf{p}_{0}$ is the starting age-specific population vector,
and $\mathbf{A}$ is the matrix. This single-sex version of the matrix
has the format shown in \eqref{eq:lesliemat-1}, with $_{5}S_{a}$
representing five year survival at age $a$ and $_{5}F_{a}$ representing
the five year fertility rate at age $a$; zeros and elided entries
are ``structural zeros'' -- elements which must always be zero in
the model (these include fertility to women in non-childbearing years,
and transitions besides a single forecast step).
\par\end{flushleft}

\begin{equation}
\begin{bmatrix}_{5}p_{0}^{0}\\
_{5}p_{5}^{0}\\
_{5}p_{10}^{0}\\
_{5}p_{15}^{0}\\
\vdots\\
_{5}p_{80}^{0}\\
_{+}p_{85}^{0}
\end{bmatrix}\label{eq:popvector-1}
\end{equation}
\begin{eqnarray}
\left[\begin{array}{cccccccc}
0 & 0 & _{5}F_{10} & _{5}F_{15} & \ldots & 0 & 0 & 0\\
_{5}S_{0} &  &  &  & \ldots\\
 & _{5}S_{5} &  &  & \ldots\\
 &  & _{5}S_{10} &  & \ldots\\
 &  &  & _{5}S_{15} & \ldots\\
 &  &  &  & \ddots\\
 &  &  &  & \cdots & _{5}S_{80}\\
 &  &  &  & \ldots &  & _{+}S_{85} & 0
\end{array}\right]\label{eq:lesliemat-1}
\end{eqnarray}

\subsubsection{Two sex Leslie matrix with migration}

\begin{flushleft}
The transition matrix generated by our version of WM follows the approach
of the work of Rogers in two-sex transition matrices \cite{rogers1},
though it is a new formulation that uses the diagonal to account for
net migration. Our matrix format is 36x36, which can be thought of
as four 18x18 blocks. The upper left block projects males, with the
sub-diagonal in this block representing male survival and the diagonal
representing male net migration. The lower right block projects females,
with the same structure as the upper left block, except that it also
contains entries in the top row of the block which generate female
babies. The upper right block's top row generates male babies; it
is placed in this block because it represents a transition from female
mothers to male children. This matrix does not model a population
closed to migration, so the rows can sum to any value, including negative
numbers. See \eqref{eq:lesliemat} for a schematic, with ``F'' representing
fertility, ``S'' survival, ``M'' net migration, and the zeros
standing for structural zeros in the matrix. ``MF'' represents ``migratory
fertility,'' babies born to mothers migrating in to the population.
To project a population forward, one simply multiplies this matrix
and a 36 x 1 population vector, as $\mathbf{p}_{1}=\mathbf{A}\mathbf{p}_{0}$.
\par\end{flushleft}

\begin{flushleft}
This form of the transition matrix is comprised of fairly standard
pieces, but it differs from other Leslie matrix formulations in important
ways. The upper left corner elements in the lower right and upper
right blocks ($MF_{m}$ and $MF_{f}$, respectively) encode fertility
dynamics, babies born in transit to mothers migrating into the population;
the lower right corner in the upper left and lower right blocks encodes
both migration and survival of the 85+ open interval. The rest of
the diagonal cells describe net migration -- population change unaccounted
for by cohort survival, the residual after the population has aged
forward. By placing migration in the diagonal and survival in the
sub-diagonal, the transition matrix is more easily interpreted in
terms of demographic components of change. However, for counties that
have high college populations, both the sub-diagonal and the diagonal
play a role in migration (see the discussion on the sub-diagonal constraint
below for more detail). Placing migration in the diagonal assumes
that the population “at risk for net migration” is the current age
group, corresponding to the method for calculating migration in \cite{johnsonmig}.
However, in the literature, the population at risk for migration is
more often considered to be the age class younger than the new migrants,
and this net migration would be accounted for in the sub-diagonal
as a cohort change ratio, in contrast to our formulation. Caswell
uses Leslie matrices primarily in the context of stage classified
models and closed populations; he uses the diagonal for those individuals
remaining in a given stage, but requires that a given row sum to less
than unity because, by definition, there is no migration in a closed
population. None of these approaches uses the corner elements for
migratory fertility, but we found their use to yield better fits through
experimentation. Our matrix formulation, though slightly original,
works well for its purpose.
\par\end{flushleft}

\begin{eqnarray}
\left[\begin{array}{cccccc|cccccc}
0 & 0 &  & \ldots &  &  & MF_{m} & 0 & _{5}F_{10} & \ldots & 0 & 0\\
_{5}S_{0} & M_{5} &  & \ldots &  &  &  &  &  & \ldots\\
 & _{5}S_{5} & M_{10} & \ldots &  &  &  &  &  & \ldots\\
 &  &  & \ddots &  &  &  &  &  & \ldots\\
 &  &  & \ldots & _{+}S_{85} & _{+}M_{85} &  &  &  & \ldots\\
\hline  &  &  &  &  &  & MF_{f} & 0 & _{5}F_{10} & \ldots & 0 & 0\\
 &  &  &  &  &  & _{5}S_{5} & M_{5} &  & \ldots\\
 &  &  &  &  &  &  & _{5}S_{5} & M_{5} & \cdots\\
 &  &  &  &  &  &  &  &  & \ddots\\
 &  &  &  &  &  &  &  &  & \ldots & _{+}S_{85} & _{+}M_{85}
\end{array}\right]\label{eq:lesliemat}
\end{eqnarray}

\subsection{Wood's method}

\begin{flushleft}
We will now describe the optimization algorithm at the core of the
technique we are presenting. Caswell (p144) \cite{mpm} gives a derivation
and an extended example of the technique using a simple 2 by 2 transition
matrix, but his exposition combines mathematical proof and algorithmic
description and can be hard to follow. Our exposition will draw very
closely from that excellent work, but will tend to a more algorithmic
approach and will not attempt to justify the method rigorously. Note
that if the reader compares the approach presented here to Caswell's,
they will see that we use a completely different constraint set, and
that Caswell uses an obsolete version of Matlab's quadratic optimization
function\texttt{ qp()}.%
\footnote{\begin{lyxlist}{00.00.0000}
\item [{Matlab}] now uses \texttt{quadprog()}, while we use Octave's \texttt{qp(),}
which has a different parameter order than Matlab's original function.
Octave's implementation of \texttt{qp() }uses a null space active
set method. In \cite{sprague-math}, we will explore the implications
of various quadratic optimization methods.\end{lyxlist}
}
\par\end{flushleft}

\begin{flushleft}
Quadratic optimization is typically formulated as follows:
\par\end{flushleft}

\begin{flushleft}
\begin{eqnarray}
minimize &  & \frac{1}{2}\mathbf{p}^{\mathbf{T}}\mathbf{Q}\mathbf{p}-\mathbf{\mathbf{c^{T}}p}\label{eq:quadraticprog}\\
subject\; to &  & \mathbf{Gp}\leqq\mathbf{h}\label{eq:Ineqconst}\\
 &  & \mathbf{Ap}=\mathbf{b}\label{eq:Eqconst}
\end{eqnarray}
\textbf{Q} is a constant matrix, $\mathbf{c}$ is a constant vector,
and \textbf{p} is the decision variable (a vector) that is varied
to find the minimal value of \eqref{eq:quadraticprog}. The matrix
\textbf{G} and vector $\mathbf{h}$ encode the inequality constraints
on \textbf{p}, and the matrix \textbf{A} and vector $\mathbf{b}$
encode the equality constraints (see above for a short introduction
to constrained optimization). The insight of Wood's method is to use
empirical population count vectors to create the matrix \textbf{Q
}and to use \textbf{p} to hold the non-zero elements of the Leslie
matrix, so that these elements of \textbf{p} can be optimized to effect
the best possible transition between age/ sex specific population
counts encoded in $\mathbf{Q}$. We will describe this in some depth.
\par\end{flushleft}

\begin{flushleft}
Note that this method optimizes the transitions between adjacent input
periods pairwise, but it does not optimize the transition from the
first period of input to data to the last. Our approach has the advantage
that it is a linear sum of sqares fitting problem$\:\left\Vert \mathbf{n}(t+1)-A\mathbf{n}(t)\right\Vert ^{2}$,
so it is extremely ``well behaved'' analytically. It is also especially
appropriate for the sort of short term forecasts that are regularly
encountered in applied work at the state and local levels. The disadvantage
is that the method may not be suitable for long term forecasting,
especially when rates are changing as well as population numbers;
in these situations one would want to optimize$\:\left\Vert \mathbf{n}(t)-A^{t}\mathbf{n}(0)\right\Vert ^{2}$,
the best fit of a single matrix for the entire input. However this
latter objective function yields nonlinearities and is likely not
as appropriate to short term forecasts as Wood's Method.%
\footnote{This insight and formulation are due to Prof. Ken Wachter at UC Berkeley.%
}
\par\end{flushleft}

\begin{flushleft}
Let \textbf{$\mathbf{n}(t)$} be $t+1$ 36 x 1 age/ sex population
vectors, indexed by time over $t=0\ldots T$, where $t$ represents
a single 5 year time step, as in (\ref{eq:popvector-1}). 
\par\end{flushleft}

\begin{flushleft}
Calculate $\mathbf{N}(t)=\mathbf{n}(t)\otimes\mathbf{I}_{36}$, where
$\otimes\;$ is the Kronecker product, and $\mathbf{I}_{36}$ is the
36 x 36 identity matrix, for $t=0\ldots T-1$. In (\ref{eq:Nkron}),
we show an example of $\mathbf{N}(t)$, in which we represent males
age zero to five as $_{5}m_{0}$ and females age 85+ as $_{+}f_{18}$
(85 plus being the 18th age class), etc. 
\par\end{flushleft}

\begin{flushleft}
\begin{equation}
\mathbf{N}(t)=\left[\begin{array}{ccccccccccccc}
_{5}m_{0}\text{0} & 0 & \, & 0 & _{5}m_{5} & 0 & \, & 0 & 0 & _{+}f_{85} & 0 & \, & 0\\
0 & _{5}m_{0} & \, & 0 & 0 & _{5}m_{5} & \, & 0 & 0 & 0 & _{+}f_{85} & \, & 0\\
\, & \, & \ddots & \, & \, & \, & \ddots & \, & \, & \, & \, & \ddots & \,\\
0 & 0 & \, & 0 & 0 & 0 & \, & _{5}f_{80} & 0 & 0 & 0 & \, & 0\\
0 & 0 & \, & _{5}m_{0} & 0 & 0 & \, & 0 & _{5}f_{80} & 0 & 0 & \, & _{+}f_{85}
\end{array}\right]\label{eq:Nkron}
\end{equation}

\par\end{flushleft}

\begin{flushleft}
Calculate $\mathbf{M}$ by vertically concatenating $\mathbf{N}(0\ldots T-1)$,
as in \eqref{eq:zstacked-1}:
\par\end{flushleft}

\begin{flushleft}
\begin{equation}
\mathbf{M}=\begin{bmatrix}\mathbf{N}(0)\\
\mathbf{N}(1)\\
\vdots\\
\mathbf{N}(T-1)
\end{bmatrix}\label{eq:zstacked-1}
\end{equation}
 
\par\end{flushleft}

\begin{flushleft}
Calculate $\mathbf{z}$ by vertically concatenating the $\mathbf{n}(1\ldots T)$
vectors, as in \eqref{eq:zstacked}:
\par\end{flushleft}

\begin{flushleft}
\begin{equation}
\mathbf{z}=\begin{bmatrix}\mathbf{n}(1)\\
\mathbf{n}(2)\\
\vdots\\
\mathbf{n}(T)
\end{bmatrix}\label{eq:zstacked}
\end{equation}

\par\end{flushleft}

\begin{flushleft}
Now let $\mathbf{p}$ be the decision vector which will contain the
elements of the fitted Leslie matrix that are not ``structural zeros''
(see \ref{eq:lesliemat} for a schematic). We seek to minimize $\left\Vert \mathbf{z-Mp}\right\Vert ^{2}$,
the distance between the projected populations formed by $\mathbf{Mp}$
and the empirical populations stored in $\mathbf{z}$.
\par\end{flushleft}

\begin{eqnarray}
\left\Vert \mathbf{z-Mp}\right\Vert  & = & (\mathbf{z-Mp)^{T}(\mathbf{z-Mp)}}\\
 & = & \mathbf{z^{\mathbf{T}}z}-\mathbf{z^{T}M^{T}p}-\mathbf{p^{T}M^{T}z+\mathbf{p^{T}M^{T}Mp}}
\end{eqnarray}

\begin{flushleft}
We drop the $\mathbf{z^{T}z}$ term, since it is constant and thus
won't play a role in the optimization, and we collect like terms to
yield:
\par\end{flushleft}

\begin{flushleft}
\begin{equation}
min\;\frac{\mathbf{p^{\mathbf{T}}M^{T}Mp}}{2}-z^{T}\mathbf{Mp}
\end{equation}

\par\end{flushleft}

\begin{flushleft}
Because is $\mathbf{M^{T}M}$ is positive definite, this objective
function is convex. Since the constraints are all linear, the optimization
problem is well defined, and we are guaranteed to find a unique global
minimum as long as the problem is feasible. The matrices $\mathbf{\frac{M^{T}M}{2}}$
and $\mathbf{z}^{T}\mathbf{M}$ are both passed into \texttt{qp()}
as parameters. 
\par\end{flushleft}

\subsection{Constraints on demographic rates }

The values in the fitted Leslie transition matrix are enforced by
constraints passed to the quadratic optimization routine \texttt{qp()}.
In the discussion below $L(r,c)$ will refer to the cell in the Leslie
matrix at row = $\mbox{r}$ and column = $c$.

\begin{flushleft}
Fertility (the cells labeled ``F'' in in \eqref{eq:lesliemat})
is constrained such that the sum of these cells falls between 1.0
and 6.0. The constraint determining the sum of the fertility cells
is stored in $\mathbf{G}$ and $\mathbf{h}$ in \eqref{eq:Ineqconst}.
Fertility is also constrained such that the relative proportions of
all the fertility cells are constant, even though the total of these
cells is determined by the optimization fitting routine. This constraint
models the fact that human fertility is somewhat consistently distributed
over fertile ages in an approximately log-normal shape. These percentages
are derived from the Human Fertility Database \cite{HFD} for women
in 1980 USA, using 51.14\% as the male sex ratio. This proportional
constraint for fertility is described in Table \ref{tab:Fertility-constraints}.
The equality constraints determining fertility shape are stored in
the matrix \textbf{A} and the vector $\mathbf{b}$ in \eqref{eq:Eqconst}.
\par\end{flushleft}

\begin{flushleft}
The sum of the fertility cells is critical to the functioning of the
method, but the distribution of the fertility over ages has very little
effect on the suitability of the matrix for single period projections.
We choose the distribution described because it is convenient and
reflects an average distribution, but we are not arguing here that
this particular fertility model is a particularly good choice or not.
However, it does allow the matrix to be used for longer term projections,
when new generations pass through these fertility cells (see discussion
above for the dangers of using WM matrices for multi-period forecasting).
\par\end{flushleft}

\begin{figure}[H]
\includegraphics[width=1\columnwidth]{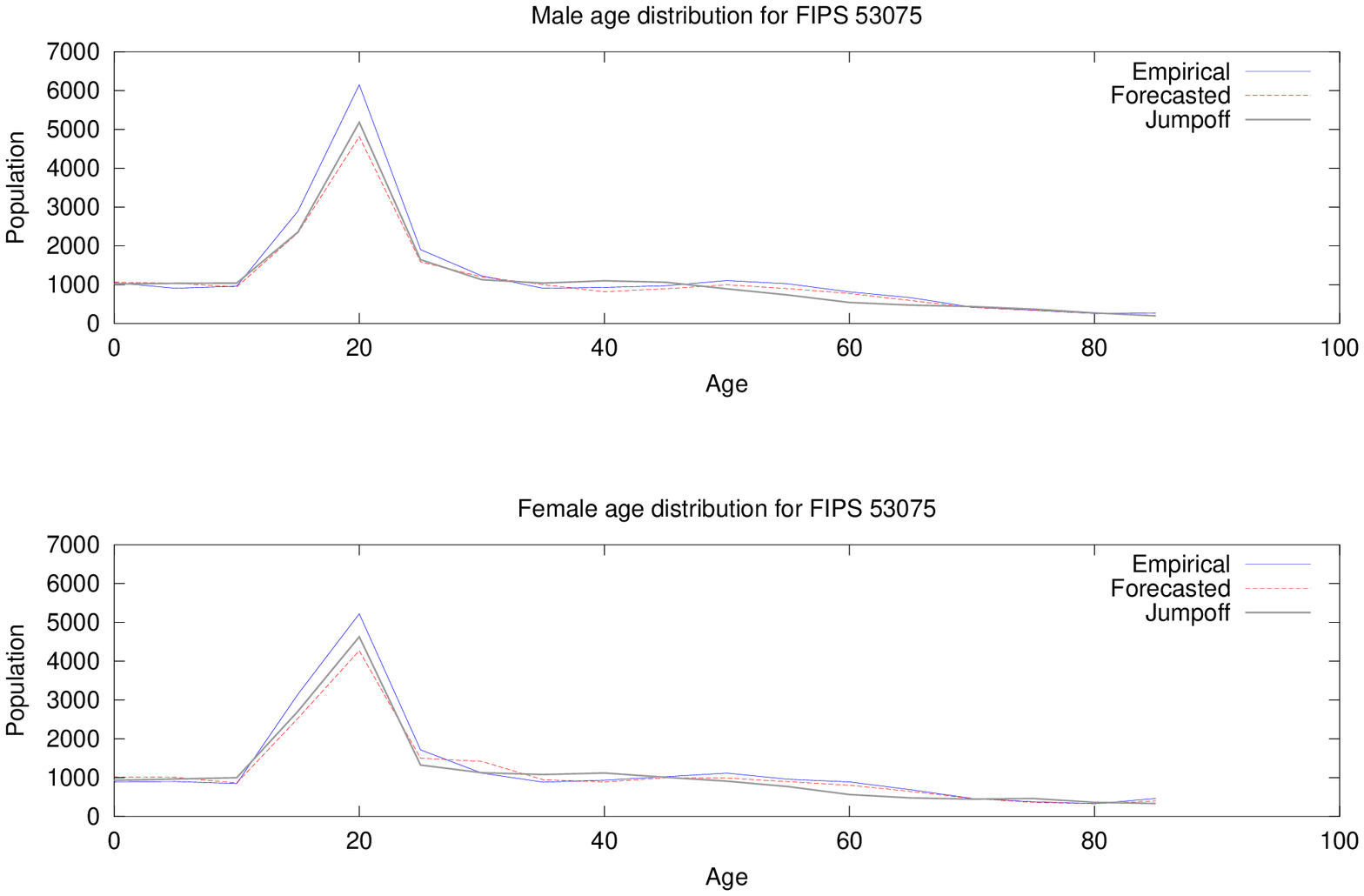}\caption{\label{fig:Whitman-County}Whitman County}

\end{figure}

\begin{flushleft}
Survival (``S'' in \eqref{eq:lesliemat}) is constrained to fit
between a lower bound corresponding to 1970 USA males and an upper
bound corresponding to 2008 USA females, computing survival as $S=Lx_{i}/Lx_{i-1}$.
Each cell in the subdiagonal is also constrained to be less than or
equal to the next younger cell. This approach fails to capture relationships
between ages (such as a mortality shape), but when these relationships
are better specified, they can be easily incorporated as constraints.
However, if the cohort ratio for 15 to 20 year olds is above 1.4,
the sub-diagonal constraints for this age are relaxed and allowed
to range between 0.4 and 1.0. This change in constraints is made because
counties with this attribute typically contain large university populations,
and without relaxing the constraints on the sub-diagonal the population
of 20-25 year olds will ``age forward'' instead of migrating out,
and an inaccurate population wave will propagate through the county
age structure (see Figure \ref{fig:Whitman-County} for a plot of
the age structure of a university-dominated county). These inequality
constraints on survival cells are described in Table \ref{tab:Survival-constraints},
and stored in matrix \textbf{G} described above.
\par\end{flushleft}

\begin{flushleft}
Migration (``M'' in \eqref{eq:lesliemat}) is constrained based
on the age -- it is allowed to range widely for people in their early
twenties and retirement ages, but it is kept smaller for other ages.
The corner cells ($L(1,19)$ and $L(19,19)$ -- for 0-5 males and
0-5 females respectively) are considered part of the fertility row
and are constrained to be part of the fertility proportion described
above, because the solver otherwise puts all the fertility dynamics
into these cells. The entry for 85+ year-olds ($L(18,18)$ and $L(36,36)$
for males and females, respectively) does “double duty” for mortality
and migration.  These inequality constraints on the migration cells
are listed in Table \ref{tab:Diagonal-migration-constraints}, and
stored in matrix \textbf{G} described above. 
\par\end{flushleft}

\begin{flushleft}
In this approach, migration is calculated as a residual; the bounds
on fertility and survival are fairly tight and the remaining dynamics
are forced into the migration cells. However, if one constrained the
migration and fertility based on known rates, the sub-diagonal would
give an estimate of survival.
\par\end{flushleft}

\begin{flushleft}
These constraints are encoded in matrices and passed to the Octave
\texttt{qp()} function along with the formulations in the above section.
Due to the large number of constraints (over 100), we don't show their
matrix formulations.
\par\end{flushleft}

\section{Optimization and demography}

\begin{flushleft}
We believe that the approach of WM taken here -- determining the elements
of a transition matrix based on an optimal fit to data, using prior
theoretical knowledge to constrain the elements and impose (linear)
relationships between them -- signals a new approach to demographic
modeling which has the potential to yield many fruitful results beyond
age/ sex forecasting. Constraints allow almost arbitrary prior knowledge
to be easily and incrementally incorporated into the inference stage
of the forecast. Using a transition matrix to store the population
dynamics gives us a succinct, interpretable representation of population
change. Using optimization to find a best fit transition matrix allows
us to go beyond deterministic modeling using fixed input rates. The
constrained optimization approach draws on a wealth of prior research
into constrained optimization and mathematical programming; working
within this well established framework may enable fruitful collaborations.
Finally, this basic approach is not limited to age/ sex forecasting,
but is applicable to any situation that can be modeled with transition
matrices. (A similar flexibility and power is possible in Bayesian
network approaches, but these only answer questions about probability
distributions; admittedly systems of probability distributions encompass
a huge number of models, but they can't model population processes
directly.) 
\par\end{flushleft}

\begin{flushleft}
Given these large claims, and the fact that in operations research
it is well known that constraints are where ``the magic happens,''
it is worth exploring our choice of constraints above in light of
traditional demographic ideas. 
\par\end{flushleft}

\begin{flushleft}
First, we think of constraints as being ``another level of indirection''
above model demographic rates. The constraints used in this paper
are based on model rates from large scale demographic data collection
projects, specifically the Human Mortality Database \cite{HMD} and
the Human Fertility Database \cite{HFD}, and from demographic theory
regarding invariants in the age structure of mortality and fertility.
However, rather than use a set of model rates deterministically, the
constrained optimization procedure allows the population count data
to have an influence on the fitted matrix within upper and lower bounds
determined by the model rates. On the other hand, if an analyst wants
to fix demographic rates deterministically, this is trivially accomplished
by setting equality constraints rather than the inequality constraints
presented here. 
\par\end{flushleft}

\begin{flushleft}
We also find it useful to think in terms of a few ``constraint sets''
rather than in terms of atomistic constraints -- we use a lower-bound
constraint set for survival, an upper-bound constraint set for survival,
an equality constraint set to enforce fertility proportions, etc.
Each constraint set is almost directly analogous to a model rate profile,
except that it provides a bound or a relationship rather than a set
of deterministic values. A constraint set may be derived empirically,
analogously to ``borrowing'' a lifetable to model a population for
which life tables have not been tabulated (often due to small sample
problems). Constraint sets can also be derived using the spine-plus-parameter
ideas of Brass \cite{wachter-textbook}, from a purely analytic expression
like the Gompertz mortality model, or using the quadratic model of
\cite{wilmoth-quadratic}. A forecaster must still decide on the values
in a constraint set -- a mortality constraint set for high mortality
countries would be derived with a higher $\alpha$ in the Gompertz
approach, for example -- but the additional level of indirection allows
for a combination of analyst judgement, information ``borrowing,''
and data inference that is impossible with deterministic model rates
alone. Finally, by using generic constraint sets with fairly wide
bounds, the analyst can create reasonable forecasts with little population-specific
knowledge (the approach taken in this paper). In fact, the analyst
can get decent forecasts with the defaults presented here, even with
no demographic theoretical background.
\par\end{flushleft}

\begin{flushleft}
Additionally, the constraint approach allows us to gracefully and
incrementally combine small- and large- scale data sources, by modifying
constraint sets with ``point constraints.'' Establishing rates for
small populations is a chronic problem in demography, especially for
mortality due to its rareness, so demographers typically ``borrow''
rates from large population lifetables when forecasting small populations.
In the constraint approach, however, one can use local knowledge to
adjust these general rates by adding point constraints at certain
ages -- for example, if homicide among 20 to 25 year old males is
a known issue in a specific population, one can force survival at
this age to be arbitrarily low without changing any other elements
of the mortality constraint set, and then fit the transition matrix
using these modified assumptions. Adjustments to the survival constraints
for university dominated counties in this paper is another example
of local, small-data information being included for specific populations
at the constraint level. 
\par\end{flushleft}

\begin{flushleft}
Finally, the constraint approach is not limited to bounding single
matrix elements. Constraints can also be written as arbitrary linear
combinations of the Leslie matrix elements, which allows for staggeringly
complex interrelationships to be enforced. This technique is used
with the fertility elements of the matrix: we enforce a constraint
such that the sum of fertility elements is between 1.0 and 6.0, and
that the fertility elements must have a fixed proportion to each other.
As a more speculative example, we might constrain migration at ages
20 to 30 to be a proportion of retirement migration (as the former
ages may work in service industries for the latter) and examine the
resulting matrix for plausibility or theoretical interest.
\par\end{flushleft}

\begin{flushleft}
More extensive analysis of this framework, including optimization
duals and approaches to confidence intervals and sensitivity analyses,
will be presented in \cite{sprague-math}. However, it is important
to understand the power of this approach and its relationship to traditional
demographic methods.
\par\end{flushleft}

\section{Testing }

\subsection{Data -- US Counties}

\begin{flushleft}
The test data is an almost complete set of U.S.A. county populations
from the US Census censal data and intercensal estimates, from 1970
through 2010, at 5 year intervals \cite{census2010}. The populations
at each interval are divided into 5 year age widths, for males and
females separately. The data was compiled from a variety of locations
on the US Census website, often aggregating over race detail in the
original data. Then the dataset was analyzed to determine which counties
maintained a consistent FIPS code over the 40 years, and only those
counties whose code was consistent were retained. This final dataset
was stored as a 36 x 3120 x 9 three dimensional matrix, with 36 age/sex
rows, 3120 county columns, and a third dimension storing the 40 years
of data. 
\par\end{flushleft}

\begin{flushleft}
Below we will present the results of this test, but in \cite{sprague-projections}
we will explore comparisons between our forecasts and other official
forecasts, perform sensitivity tests, and examine the error structure
somewhat exhaustively. In that paper, we will also use selected international
data, and consider other geographies in the USA besides counties.
Additionally, we will examine demographic rates (births, deaths, and
migration) implied by the Leslie matrix projection. We have chosen
to separate the exhaustive empirical paper from this paper in order
to keep the exposition clear.
\par\end{flushleft}

\begin{flushleft}
To test the method, each county was run separately to compare forecasted
population with actual population. The age specific population in
1980 through 2000 was used as the training set. This set was processed
by an Octave routine that reads in population data and default constraints
and returns the transition matrix using the method outlined above.
These resulting transition matrices were used to forecast 2000 population
counts two steps forward, yielding an age specific forecast for 2010
using simple matrix multiplication $\mathbf{p}_{t+2}=\mathbf{A}^{2}\mathbf{p}_{t}$.
The resulting population vector for 2010 was compared to the empirical
data for 2010, and absolute percentage differences were computed for
each age and sex cell. These percentage differences and other metadata
were stored for all the counties as another set of 36 by 3120 matrices.
\par\end{flushleft}

\begin{flushleft}
There are some caveats with this dataset, though we believe it provides
a sufficient test of the method to prove its utility. Two counties
failed to converge when their data was processed through the Woods
method, and these were dropped from the error analysis below. Certain
important counties were dropped because they had changes in their
FIPS codes over the input period, including Dade County in Florida.
Counties which changed their geographic boundaries but retained their
FIPS code are assumed to be homogeneous across time, even though their
populations could have changed in ways that might bias the fitted
Leslie matrix. We used intercensal estimates for the periods in between
censuses (1975, 1985, 1995, 2005). Finally, data quality is at the
mercy of the US Census Bureau.
\par\end{flushleft}

\begin{flushleft}
Counties were chosen as a unit of analysis for several reasons. First,
this research was initiated to support Growth Management Act (GMA)
forecasting at Washington State, and counties are the basic unit as
determined by GMA legislation. Second, institutions throughout the
United States are regularly called upon to forecast population at
the county level, so these techniques would frequently be exercised
on county data if they were adopted. Third, county data is of excellent
quality overall, as counties form both a basic enumeration and a basic
tabulation geography for the U. S. Census Bureau. Fourth, counties
provide a wide range of population sizes on which to test the method,
ranging in 2010 population from 9,818,605 (Los Angeles County, California)
to 82 (Loving County, Texas). Finally, counties also show a diversity
of of population dynamics due to their sometimes very particular social
contexts, with smaller counties often ``specializing'' in ways that
influence age structure, such as providing retirement communities
or housing university student populations. 
\par\end{flushleft}

\begin{flushleft}
This dataset should be considered as an extended example of the method
but not an exhaustive analysis of the method's error properties.
\par\end{flushleft}

\subsection{Error structure of ten year forecasts}

\begin{flushleft}
After all the counties were run, the 2010 forecasts were compared
to actual 2010 census counts.
\par\end{flushleft}

\begin{flushleft}
The Mean Absolute Percentage Error (MAPE) over all cells is 10\%;
this is comparable to the MAPEs for 10 year county forecasts given
in Smith and Tayman \cite{smithfcsts,smithtaymanfcsts}. If we test
the method on counties with populations larger than 50,000, the MAPE
over all cells is only 7\%, and the MAPE for counties with less than
50,000 population is 11\%. The age specific MAPEs are slightly larger,
see Table \ref{tab:MAPE-by-age}. Here we see that the age specific
MAPEs are comparable to the forecasts presented in Smith and Tayman
(p 749) for several county forecasts in Florida. It is hard to compare
them because Smith and Tayman use ten year age widths in their evaluation
forecasts and forecast errors are generally smaller with large populations
and/ or wider age groups. Smith and Tayman also only use county forecasts
for the state of Florida, while we produced forecasts over almost
all counties in the US. (Unfortunately the US Census Bureau does not
produce age/ sex population projections for counties, so it is difficult
to assemble a large dataset against which to compare this method at
the county level.)Also interesting is the quantile pattern of errors
in Table \ref{tab:Quantile-Error-structure}. 
\par\end{flushleft}

\subsection{County examples}

\begin{figure}[H]
\includegraphics[width=1\columnwidth]{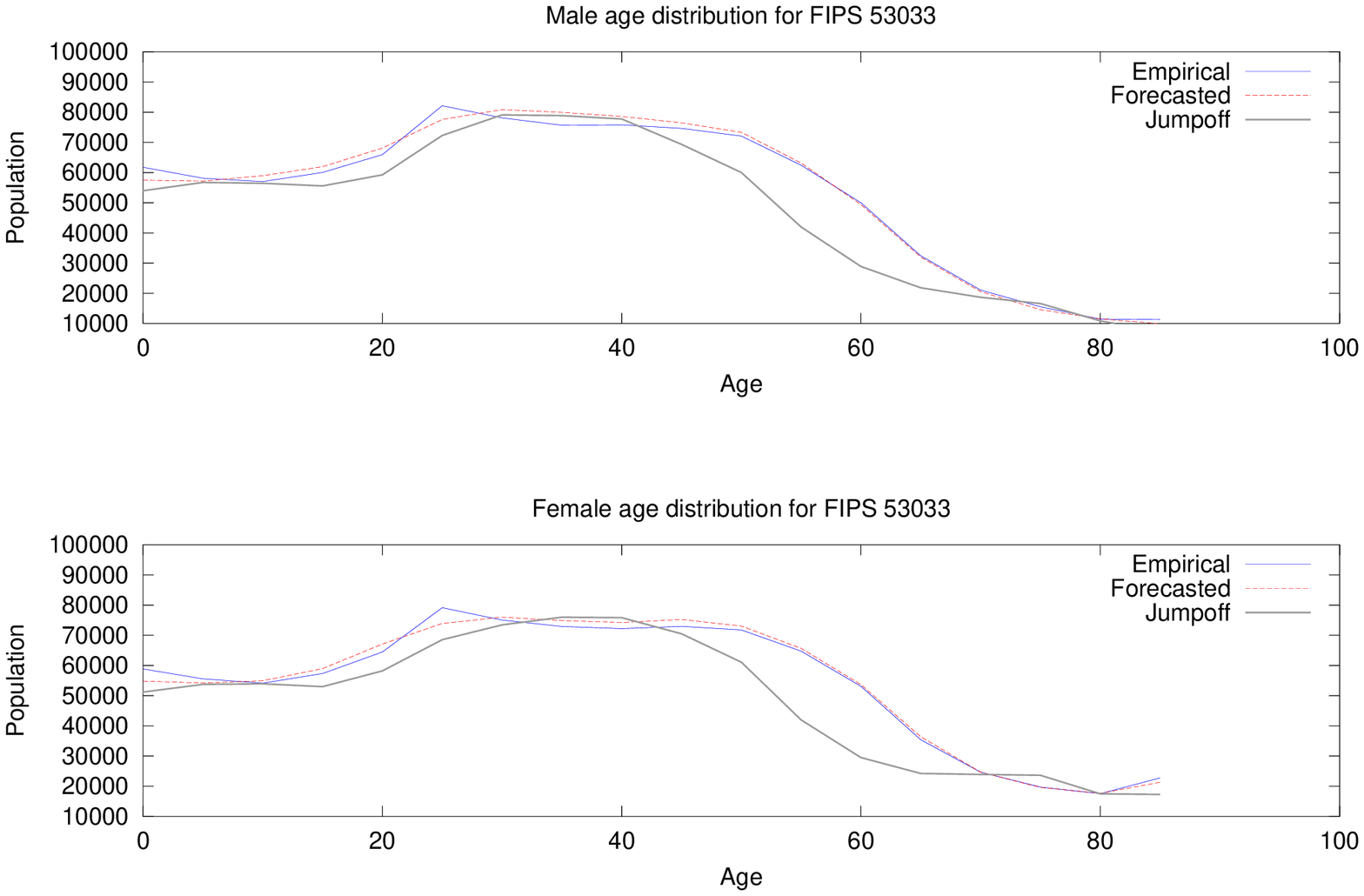}\caption{\label{fig:King-County}King County}
\end{figure}

\begin{figure}[H]
\includegraphics[width=1\columnwidth]{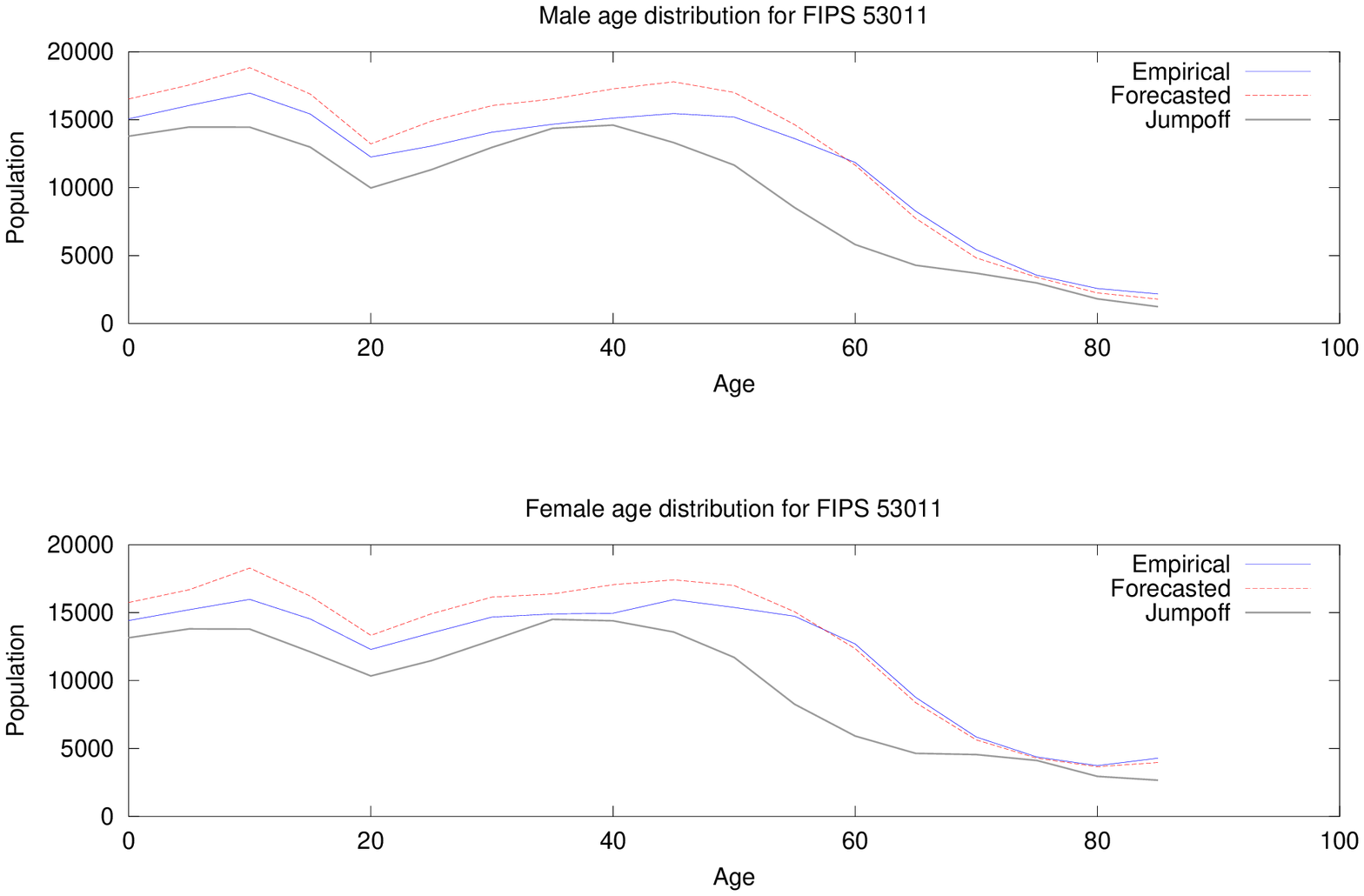}\caption{\label{fig:Clark-County}Clark County}

\end{figure}

\begin{figure}[H]
\includegraphics[width=1\columnwidth]{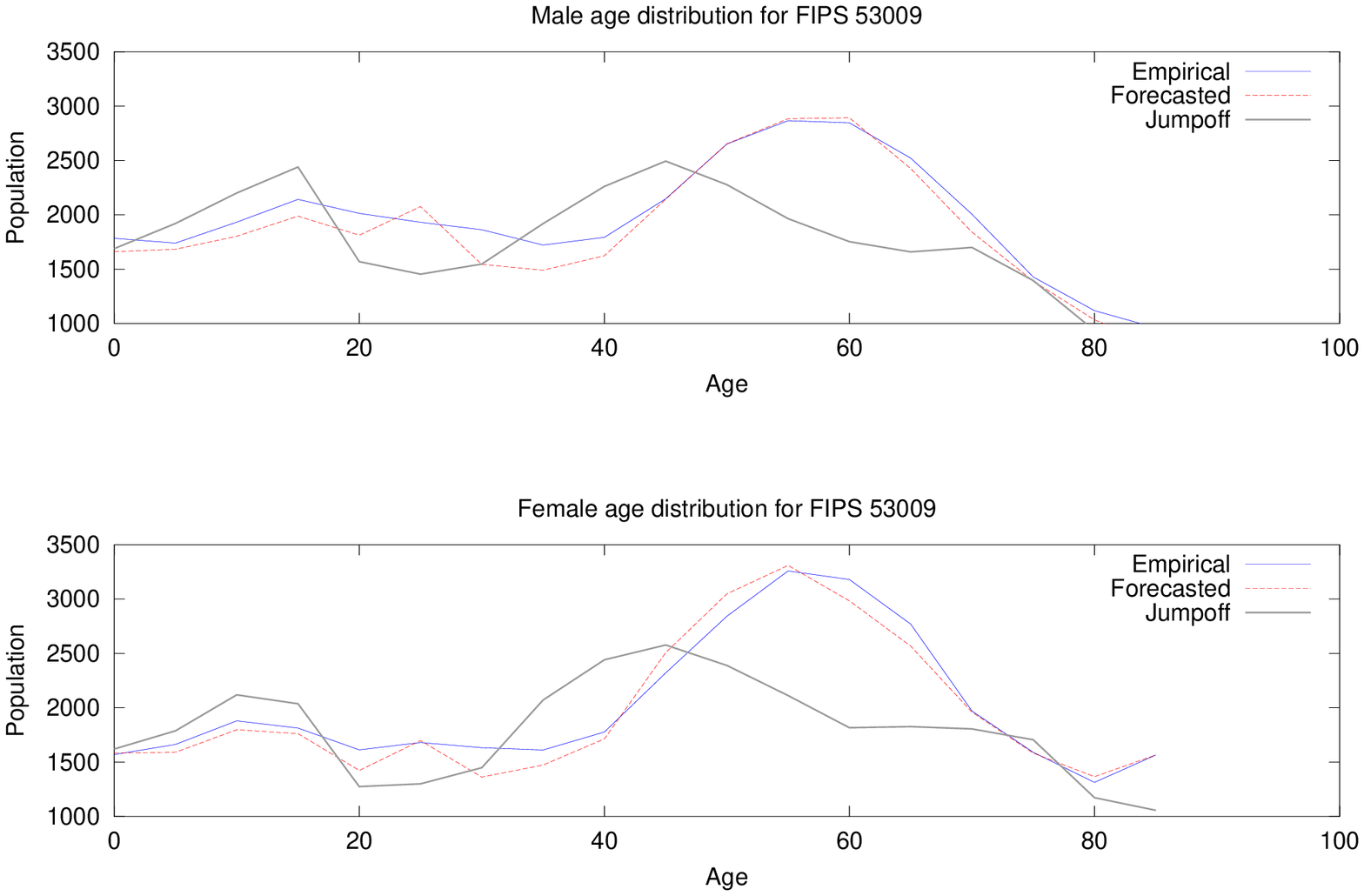}\caption{\label{fig:Clallam-County}Clallam County}

\end{figure}

\begin{figure}[H]
\includegraphics[width=1\columnwidth]{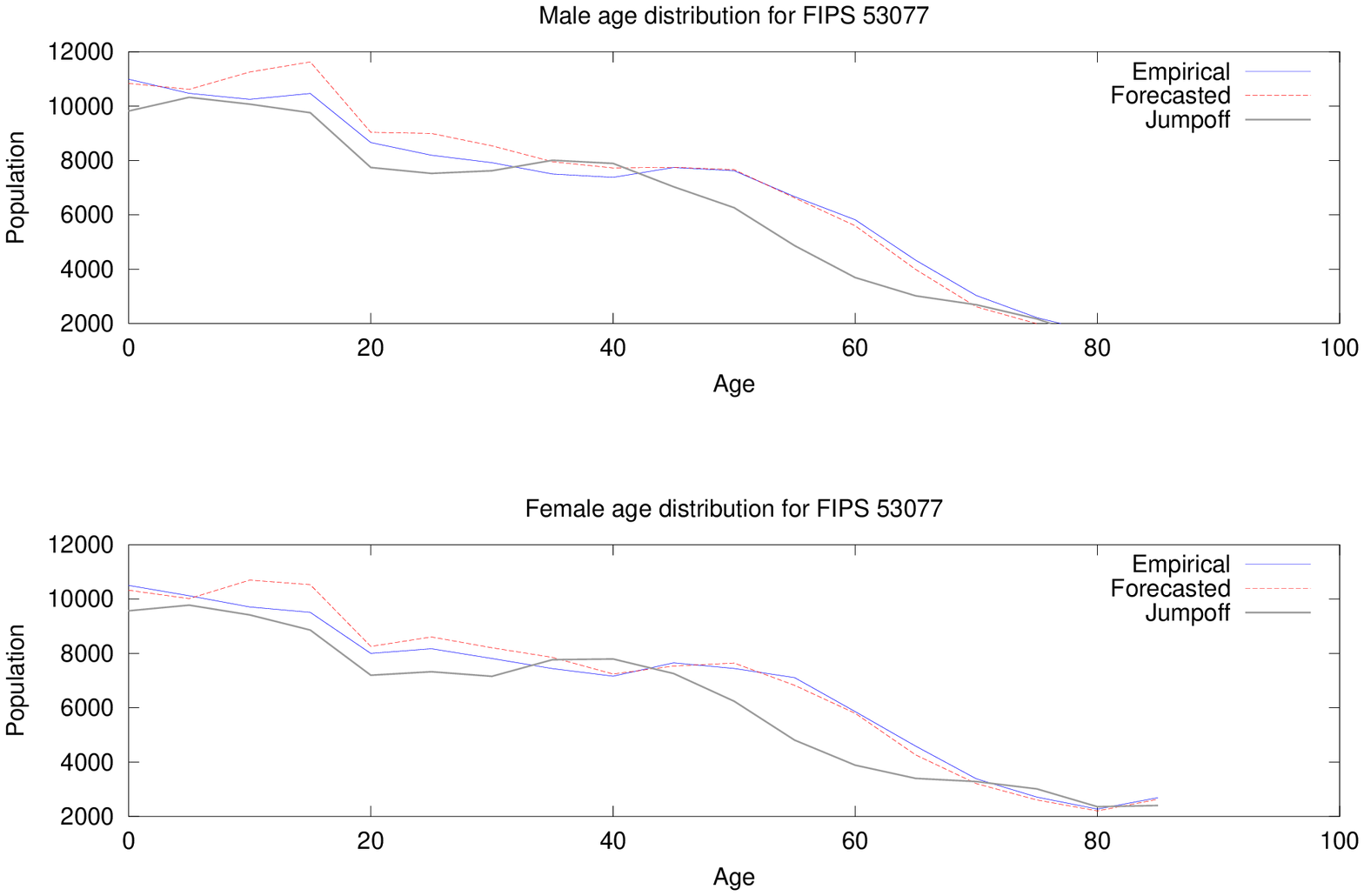}\caption{\label{fig:Yakima-County}Yakima County}

\end{figure}

\begin{flushleft}
As a further example of the utility and limitations of the method
presented, we consider several counties in Washington State (see figures
\ref{fig:Spokane-County} through \ref{fig:Yakima-County}). In the
plots, the solid blue line shows the empirical population in 2010,
the dashed red line is the 2010 forecasted population, and the gray
line is the jump off population in 2000. 
\par\end{flushleft}

\begin{flushleft}
These plots allow us to see many of the basic population dynamics
for each county, and show that our method forecasts the age/ sex composition
well. If a peak or a trough in the gray 2000 line moves rightward,
to the blue and red 2010 lines, this implies that part of the cohort
is aging forward and staying in the county. If the peaks and troughs
don't move to the right, but rather move up and down in the same age,
it implies that people are moving into the area at a given age and
back out again when they get older (see Whitman County for an example
-- people move into the county for college and move out when they
graduate). We will use the terms ``cohort age dynamic'' and ``migratory
age dynamic'' for these two population dynamics, respectively.
\par\end{flushleft}

\begin{flushleft}
These counties were chosen to represent specializations in county
attributes and corresponding ages of residents. They include counties
with large regional centers (Spokane County), large internationally
important cities (King County, which contains Seattle), suburban counties
(Clark County, which provides housing for the Portland, OR metropolitan
area), high retirement population counties (Clallam County), university
oriented counties (Whitman County), and agricultural counties (Yakima
County); each of these specializations have distinct age structures,
all of which are forecasted well by Woods Method. Although these examples
were chosen informally based on our knowledge of their social context,
the idea of county classification owes much to \cite{Pitt,Parker-2006}. 
\par\end{flushleft}

\begin{flushleft}
Note that population growth in Washington state has slowed considerably
below long term averages since 2008 due to the recession (in-migration
was 13,000 in 2010 and about 4,000 in 2011, compared to a yearly average
of about 45,000). This slowdown affects the accuracy of the forecast
totals, since the training period had higher overall growth than the
forecast period. Even when the total forecast is lower than the empirical
number, however, Wood's method often retains the basic shape of the
age structure; this ability to keep structure is a benefit of the
method.
\par\end{flushleft}

\subsubsection{Regional city center (Spokane County, FIPS 53063)}

\begin{flushleft}
Spokane County is located on the Eastern border of Washington. It
has a population of about 470,000, a diverse and vibrant economy,
including manufacturing, farming, and multiple universities (Gonzaga,
Eastern Washington University, and other smaller private colleges).
Historically, Spokane County has had slow but steady growth. While
the county contains a typical metropolitan center (the city of Spokane),
the county is large enough to also include most of the family-oriented
suburbs serving the central city. It provides an example of a ``well-rounded''
county, with a fairly complete range of ages. In Figure \ref{fig:Spokane-County},
we can see a cohort age dynamic as the baby boomers and their children
both age forward in time. The forecast is quite good, except at ages
20-25, due possibly to increased college enrollment in the current
recessionary environment. 
\par\end{flushleft}

\subsubsection{Large metropolitan center (King County, FIPS 53033)}

\begin{flushleft}
King County contains Seattle, a thriving high-tech industry, a large
amount of urbanized area, and many universities, including the University
of Washington (see Figure \ref{fig:King-County}). Like many highly
urbanized counties, there is a preponderance of people in their twenties
and early thirties, who migrate into the county for employment and
education but move out to the suburbs as they start families; this
migratory age dynamic causes the steady bulge in the twenties and
early thirties that we see in the figure. We also see a fair number
of baby boomers experiencing cohort age dynamics, as that bulge extends
to older ages in 2010. Note the relative small numbers of children,
a proportion of the population that is subject to migratory dynamics
rather than cohort age dynamics due to the tendency of families to
move to the suburbs when their children reach school age. Note also
that the forecast under-predicts age 0-5, consonant with the anecdotal
evidence that in the last decade more parents are staying in urban
areas than in the 1990s and before. The forecast is good, except for
the unpredicted peak at age 25-30, possibly due to the technology's
sector strong performance even during the recessionary environment. 
\par\end{flushleft}

\subsubsection{Suburban satellite (Clark County, FIPS 53011) }

\begin{flushleft}
Clark County shows a typical suburban dynamic, in which young people
move out in their teens and early twenties, and family-age people
move in in their later twenties and thirties (see Figure \ref{fig:Clark-County}),
which is what we expect given that Clark County is a bedroom community
to nearby Portland and Washington County in Oregon. Note that we see
the baby boom peak experiencing cohort age dynamics, but the younger
ages evincing migratory age dynamics with no forward cohort movement.
Note also that the forecast over-predicts total population due to
the current recession, but it retains the shape of the population
extremely well. 
\par\end{flushleft}

\subsubsection{Retirement/ amenity (Clallam County, FIPS 53009) }

\begin{flushleft}
Clallam County is a rural county on the Olympic Peninsula, with high
recreational amenities (coast and mountains) but low employment opportunity.
Like many similar counties, it shows typical older in-migration, with
a huge bulge at about 55 years old. This age group displays cohort
aging, but it also shows migratory age dynamics as the peak increases
from migration (growing upward as well as moving to the right). The
trough at 20-25 years old has migratory age dynamics, presumably as
these young adults leave the rural county for opportunity and education.
The forecast predicts the older population very well, but the younger
population is under-forecast. 
\par\end{flushleft}

\subsubsection{Agricultural region (Yakima County, FIPS 53077)}

\begin{flushleft}
Yakima County has an agricultural economic base, a large hispanic
population, and high fertility. There is large out-migration in the
early twenties for non-hispanics. See Figure \ref{fig:Yakima-County}.
The forecast is quite good, except for over-predicting population
in their twenties, which may be due to the current recessionary environment
damping the force of Latino in-migration. 
\par\end{flushleft}

\subsubsection{University specialization (Whitman County, FIPS 53075)}

\begin{flushleft}
Whitman County contains Washington State University, a large public
university. Besides the university, however, the next most important
local industry is low-labor wheat farming, and there are few recreational
attractions in nearby; this configuration is consistent with the large
population spike at college ages (see Figure \ref{fig:Whitman-County}),
which sharply decreases at the next age interval (25-30) with young
people leaving shortly after graduation. Notice that the forecast
is for decreasing population from 2000, while the empirical data show
an increase in population. This mis-forecast is due to prior trends
being incorporated into the transition matrix through the training
data; from 1990 to 2000, Whitman saw an overall drop in population
at these ages, but due to the recent recessionary environment, enrollment
at WSU has probably seen higher enrollment in 2000-2010. Even with
this error in forecast magnitude, the shape of the forecast population
is correct. 
\par\end{flushleft}

\section{Discussion}

\subsection{Use in applied settings}

\begin{flushleft}
There are a few extensions to this method that would make it more
useful in applied settings. We sketch how these can be implemented
or discuss plans for future research relating to them.
\par\end{flushleft}

\subsubsection{Forecasting to a control total}

\begin{flushleft}
Often a population is forecast by projecting a total number, with
the age/sex specific proportions ``controlled'' to that total. This
approach is trivial to effect using Wood's Method as presented here.
First, forecast the age/sex specific populations with WM, then derive
age/sex proportion vectors for each forecast step by dividing each
age/sex specific vector by its sum, and finally multiply the projected
total forecast numbers by the corresponding age/sex specific proportions.
\par\end{flushleft}

\subsubsection{Forecasting vital counts}

\begin{flushleft}
It is also important to be able to forecast numbers of vital events
(births, deaths, net migration) along with population numbers. An
estimate of these events should be easy to derive by using the various
sections of the Leslie transition matrix corresponding to the vital
event. To estimate births, set the migration diagonal and the survival
sub-diagonal to zero, then multiply the age/sex population vector
by the the remaining matrix. A directly analogous technique can be
used for migration. For deaths, project the population using only
the survival sub-diagonal, then find the difference of the starting
vector and the finishing ``cohort-wise.'' These techniques all depend
on the choice of intelligent constraints in the original fitting routine,
as otherwise the matrix components cannot be assumed to have demographic
interpretations. Unfortunately, this approach is impossible with college
populations, as the sub-diagonal no longer corresponds to survival;
refining the model for college counties is ongoing. We have not tested
this technique in practice.
\par\end{flushleft}

\subsubsection{Confidence intervals}

\begin{flushleft}
Determining forecast bands is a more difficult proposition. Given
the error structure described above, one could use the percentage
corresponding to each age group for the 80th percentile as a general
guideline for high and low bands. However, there are two problems
with this approach. First, there is almost surely heterogeneity within
counties with respect to their error structure; if nothing else, larger
counties have smaller percentile errors than smaller counties. More
subtly, age specific error does not address correlation in error between
ages for a given county, and so loses any sense of the overall accuracy
of the forecast shape. 
\par\end{flushleft}

\subsubsection{Choosing input data}

\begin{flushleft}
It is also important to consider how to choose input data to derive
the Leslie transition matrices. Since Wood's Method incorporates an
average of all transitions, if past trends don't reflect future trends,
the forecasts will be incorrect. There are no hard and fast rules
for choosing training data, but there are two basic rules of thumb,
which, unfortunately, can contradict each other in any given application.
One, generally forecasters recommend going as far back for input as
the forecast goes forward in time. Two, examine past data and look
for articulation points where trends change direction, and avoid using
data from before the most recent ``elbow'' in the data.
\par\end{flushleft}

\subsection{Further research}

\begin{flushleft}
There are three main directions we see for exploring the application
of this method. The first is extending it to more complex matrix models,
including race, multi-regional models, and educational state models.
The second direction will be in creating and refining constraints,
as both limits on rates and subtle population interactions are better
approximated and modeled. Finally, confidence bands and error analysis
will be a line of development; currently the author has no systematic
procedure for generating confidence bands or alternative forecasts.
Sensitivity analyses, partly based on the wealth of optimization theory
and the approaches of Caswell (and Tulja?) will be extremely important
in this line of investigation. Additionally, the mathematical properties
of the optimization method are not well explored. Do different quadratic
optimization algorithms yield better or worse results? Can the matrix
$M$ be better characterized (currently it has a very high conditioning
number, which might affect performance)?
\par\end{flushleft}

\section{Conclusion}

We believe that the forecasting method outlined above shows great
promise for use in applied settings to forecast age and sex because
it both streamlines the creation of Leslie matrices, and it flexibly
incorporates previous demographic knowledge in the constraints. 

Additionally, the method should be generally applicable to all populations
using the above approach of combining theoretically based constraints
with empirical data. While different populations will require different
matrices and constraint sets, the basic applicability of the approach
has been shown.

\newpage{}

\section{Tables}

\begin{table}[H]
\begin{tabular}{|c|r@{\extracolsep{0pt}.}l|r@{\extracolsep{0pt}.}l|c|}
\hline 
Age & \multicolumn{2}{c|}{males} & \multicolumn{2}{c|}{females} & cells\tabularnewline
\hline 
\hline 
0 & 2&56\% & 2&44\% & $a_{1,19},a_{19,19}$\tabularnewline
\hline 
10 & 1&53\% & 1&47\% & $a_{1,21},a_{19,21}$\tabularnewline
\hline 
15 & 10&23\% & 9&77\% & $a_{1,22},a_{19,22}$\tabularnewline
\hline 
20 & 13&81\% & 13&19\% & $a_{1,23},a_{19,23}$\tabularnewline
\hline 
25 & 12&78\% & 12&21\% & $a_{1,24},a_{19,24}$\tabularnewline
\hline 
30 & 7&67\% & 7&33\% & $a_{1,25},a_{19,25}$\tabularnewline
\hline 
35 & 2&05\% & 1&95\% & $a_{1,26},a_{19,26}$\tabularnewline
\hline 
40 & 0&51\% & 0&49\% & $a_{1,27},a_{19,27}$\tabularnewline
\hline 
45 & 0&00\% & 0&00\% & $a_{1,28},a_{19,28}$\tabularnewline
\hline 
\end{tabular}

\caption{\label{tab:Fertility-constraints}Fertility constraints}
\end{table}

\newpage{}
\begin{table}[H]
\begin{tabular}{|c|c|c|}
\hline 
Age & Cells & Value\tabularnewline
\hline 
\hline 
5 & $a_{2,2},a_{20,20}$ & +/-0.50\tabularnewline
\hline 
10 & $a_{3,3},a_{21,21}$ & +/-0.50\tabularnewline
\hline 
15 & $a_{4,4},a_{22,22}$ & +/-1.50\tabularnewline
\hline 
20 & $a_{5,5},a_{23,23}$ & +/-1.50\tabularnewline
\hline 
25 & $a_{6,6},a_{24,24}$ & +/-3.50\tabularnewline
\hline 
30 & $a_{7,7},a_{25,25}$ & +/-1.50\tabularnewline
\hline 
35 & $a_{8,8},a_{26,26}$ & +/-0.75\tabularnewline
\hline 
40 & $a_{9,9},a_{27,27}$ & +/-0.50\tabularnewline
\hline 
45 & $a_{10,10},a_{28,28}$ & +/-0.50\tabularnewline
\hline 
50 & $a_{11,11},a_{29,29}$ & +/-0.50\tabularnewline
\hline 
55 & $a_{12,12},a_{30,30}$ & +/-0.50\tabularnewline
\hline 
60 & $a_{13,13},a_{31,31}$ & +/-0.75\tabularnewline
\hline 
65 & $a_{14,14},a_{32,32}$ & +/-0.75\tabularnewline
\hline 
70 & $a_{15,15},a_{33,33}$ & +/-0.75\tabularnewline
\hline 
75 & $a_{16,16},a_{34,34}$ & +/-0.50\tabularnewline
\hline 
80 & $a_{17,17},a_{35,35}$ & +/-0.50\tabularnewline
\hline 
85+ & $a_{18,18},a_{36,36}$ & +/-0.65\tabularnewline
\hline 
\end{tabular}\caption{\label{tab:Diagonal-migration-constraints}Diagonal migration constraints}
\end{table}
\newpage{}
\begin{table}[H]
\begin{tabular}{|c|c|c|c|c|c|}
\hline 
Age & Cells & Male min & Male max & Fem min & Fem max\tabularnewline
\hline 
\hline 
0 & $a_{2,1},a_{20,19}$ & 0.99614 & 0.99887 & 0.99709 & 0.99908\tabularnewline
\hline 
5 & $a_{3,2},a_{21,20}$ & 0.99769 & 0.99921 & 0.99854 & 0.99938\tabularnewline
\hline 
10 & $a_{4,3},a_{22,21}$ & 0.99523 & 0.99775 & 0.99784 & 0.99885\tabularnewline
\hline 
15 & $a_{5,4},a_{23,22}$ & 0.98988 & 0.9938 & 0.99645 & 0.99781\tabularnewline
\hline 
20 & $a_{6,5},a_{24,23}$ & 0.98952 & 0.99285 & 0.99605 & 0.99741\tabularnewline
\hline 
25 & $a_{7,6},a_{25,24}$ & 0.98955 & 0.99283 & 0.99503 & 0.9968\tabularnewline
\hline 
30 & $a_{8,7},a_{26,25}$ & 0.98676 & 0.99178 & 0.99258 & 0.99561\tabularnewline
\hline 
35 & $a_{9,8},a_{27,26}$ & 0.98066 & 0.98877 & 0.98875 & 0.99333\tabularnewline
\hline 
40 & $a_{10,9},a_{28,27}$ & 0.96986 & 0.98278 & 0.98285 & 0.98954\tabularnewline
\hline 
45 & $a_{11,10},a_{29,28}$ & 0.95364 & 0.97377 & 0.97465 & 0.98425\tabularnewline
\hline 
50 & $a_{12,11},a_{30,29}$ & 0.92778 & 0.96162 & 0.963 & 0.97767\tabularnewline
\hline 
55 & $a_{13,12},a_{31,30}$ & 0.89199 & 0.94611 & 0.94672 & 0.96668\tabularnewline
\hline 
60 & $a_{14,13},a_{32,31}$ & 0.84307 & 0.92023 & 0.92108 & 0.94846\tabularnewline
\hline 
65 & $a_{15,14},a_{33,32}$ & 0.78197 & 0.88355 & 0.87881 & 0.92151\tabularnewline
\hline 
70 & $a_{16,15},a_{34,33}$ & 0.69812 & 0.82576 & 0.80976 & 0.87797\tabularnewline
\hline 
75 & $a_{17,16},a_{35,34}$ & 0.59611 & 0.73736 & 0.71033 & 0.80798\tabularnewline
\hline 
80 & $a_{18,17},a_{36,35}$ & 0.46707 & 0.61004 & 0.57199 & 0.6961\tabularnewline
\hline 
\end{tabular}\caption{\label{tab:Survival-constraints}Survival constraints}
\end{table}
\newpage{}
\begin{table}[H]
\begin{tabular}{|c|r|r|r|r|}
\hline 
Age & MAPE WM & MAPE S\&T 1990 & MAPE S\&T, \#1 & MAPE S\&T 2000, \#2\tabularnewline
\hline 
\hline 
0 & 12\% & 9.6\% & 16.4\% & 10.3\%\tabularnewline
\hline 
5 & 11\% & \multirow{2}{*}{9.1\%} & \multirow{2}{*}{13.7\%} & \multirow{2}{*}{8.0\%}\tabularnewline
\cline{1-2} 
10 & 9\% &  &  & \tabularnewline
\hline 
15 & 9\% & \multirow{2}{*}{11.2\%} & \multirow{2}{*}{13.1\%} & \multirow{2}{*}{10.2\%}\tabularnewline
\cline{1-2} 
20 & 15\% &  &  & \tabularnewline
\hline 
25 & 16\% & \multirow{2}{*}{13.3\%} & \multirow{2}{*}{18.0\%} & \multirow{2}{*}{12.7\%}\tabularnewline
\cline{1-2} 
30 & 13\% &  &  & \tabularnewline
\hline 
35 & 11\% & \multirow{2}{*}{10.4\%} & \multirow{2}{*}{14.7\%} & \multirow{2}{*}{11.6\%}\tabularnewline
\cline{1-2} 
40 & 10\% &  &  & \tabularnewline
\hline 
45 & 8\% & \multirow{2}{*}{9.5\%} & \multirow{2}{*}{12.3\%} & \multirow{2}{*}{11.2\%}\tabularnewline
\cline{1-2} 
50 & 7\% &  &  & \tabularnewline
\hline 
55 & 7\% & \multirow{2}{*}{9.3\%} & \multirow{2}{*}{15.5\%} & \multirow{2}{*}{13.3\%}\tabularnewline
\cline{1-2} 
60 & 6\% &  &  & \tabularnewline
\hline 
65 & 7\% & \multirow{5}{*}{11.0\%} & \multirow{5}{*}{17.8\%} & \multirow{5}{*}{10.1\%}\tabularnewline
\cline{1-2} 
70 & 8\% &  &  & \tabularnewline
\cline{1-2} 
75 & 8\% &  &  & \tabularnewline
\cline{1-2} 
80 & 11\% &  &  & \tabularnewline
\cline{1-2} 
85 & 13\% &  &  & \tabularnewline
\hline 
\end{tabular}\caption{\label{tab:MAPE-by-age}MAPE by age}
\end{table}

\newpage{}
\begin{table}[H]
\begin{tabular}{|c|c|c|c|c|c|}
\hline 
Age & Sex & 50\% & 80\% & 97.5\% & Max\tabularnewline
\hline 
\hline 
0 & M & 10\% & 18\% & 37\% & 151\%\tabularnewline
\hline 
5 & M & 8\% & 16\% & 36\% & 117\%\tabularnewline
\hline 
10 & M & 6\% & 14\% & 36\% & 109\%\tabularnewline
\hline 
15 & M & 6\% & 13\% & 34\% & 185\%\tabularnewline
\hline 
20 & M & 10\% & 22\% & 61\% & 405\%\tabularnewline
\hline 
25 & M & 11\% & 25\% & 65\% & 323\%\tabularnewline
\hline 
30 & M & 10\% & 22\% & 55\% & 286\%\tabularnewline
\hline 
35 & M & 8\% & 19\% & 50\% & 221\%\tabularnewline
\hline 
40 & M & 7\% & 17\% & 48\% & 168\%\tabularnewline
\hline 
45 & M & 5\% & 12\% & 37\% & 215\%\tabularnewline
\hline 
50 & M & 5\% & 11\% & 34\% & 132\%\tabularnewline
\hline 
55 & M & 5\% & 11\% & 30\% & 96\%\tabularnewline
\hline 
60 & M & 5\% & 10\% & 25\% & 138\%\tabularnewline
\hline 
65 & M & 7\% & 12\% & 25\% & 129\%\tabularnewline
\hline 
70 & M & 9\% & 15\% & 29\% & 115\%\tabularnewline
\hline 
75 & M & 9\% & 15\% & 30\% & 200\%\tabularnewline
\hline 
80 & M & 11\% & 19\% & 36\% & 256\%\tabularnewline
\hline 
85 & M & 12\% & 21\% & 45\% & 704\%\tabularnewline
\hline 
0 & F & 10\% & 18\% & 36\% & 135\%\tabularnewline
\hline 
5 & F & 9\% & 17\% & 38\% & 202\%\tabularnewline
\hline 
10 & F & 6\% & 13\% & 36\% & 164\%\tabularnewline
\hline 
15 & F & 6\% & 13\% & 38\% & 298\%\tabularnewline
\hline 
20 & F & 10\% & 20\% & 61\% & 416\%\tabularnewline
\hline 
25 & F & 10\% & 23\% & 64\% & 217\%\tabularnewline
\hline 
30 & F & 9\% & 19\% & 46\% & 127\%\tabularnewline
\hline 
35 & F & 6\% & 15\% & 36\% & 112\%\tabularnewline
\hline 
40 & F & 6\% & 13\% & 34\% & 111\%\tabularnewline
\hline 
45 & F & 5\% & 11\% & 29\% & 113\%\tabularnewline
\hline 
50 & F & 4\% & 10\% & 28\% & 103\%\tabularnewline
\hline 
55 & F & 4\% & 9\% & 25\% & 146\%\tabularnewline
\hline 
60 & F & 4\% & 9\% & 21\% & 145\%\tabularnewline
\hline 
65 & F & 4\% & 9\% & 22\% & 162\%\tabularnewline
\hline 
70 & F & 4\% & 9\% & 22\% & 162\%\tabularnewline
\hline 
75 & F & 4\% & 10\% & 28\% & 226\%\tabularnewline
\hline 
80 & F & 5\% & 12\% & 31\% & 258\%\tabularnewline
\hline 
85 & F & 7\% & 17\% & 45\% & 342\%\tabularnewline
\hline 
\end{tabular}

\caption{\label{tab:Quantile-Error-structure}Quantile error structure}
\end{table}

\newpage{}

\section{Plots}

\begin{flushleft}
\begin{figure}[H]
\includegraphics[clip,width=1\columnwidth]{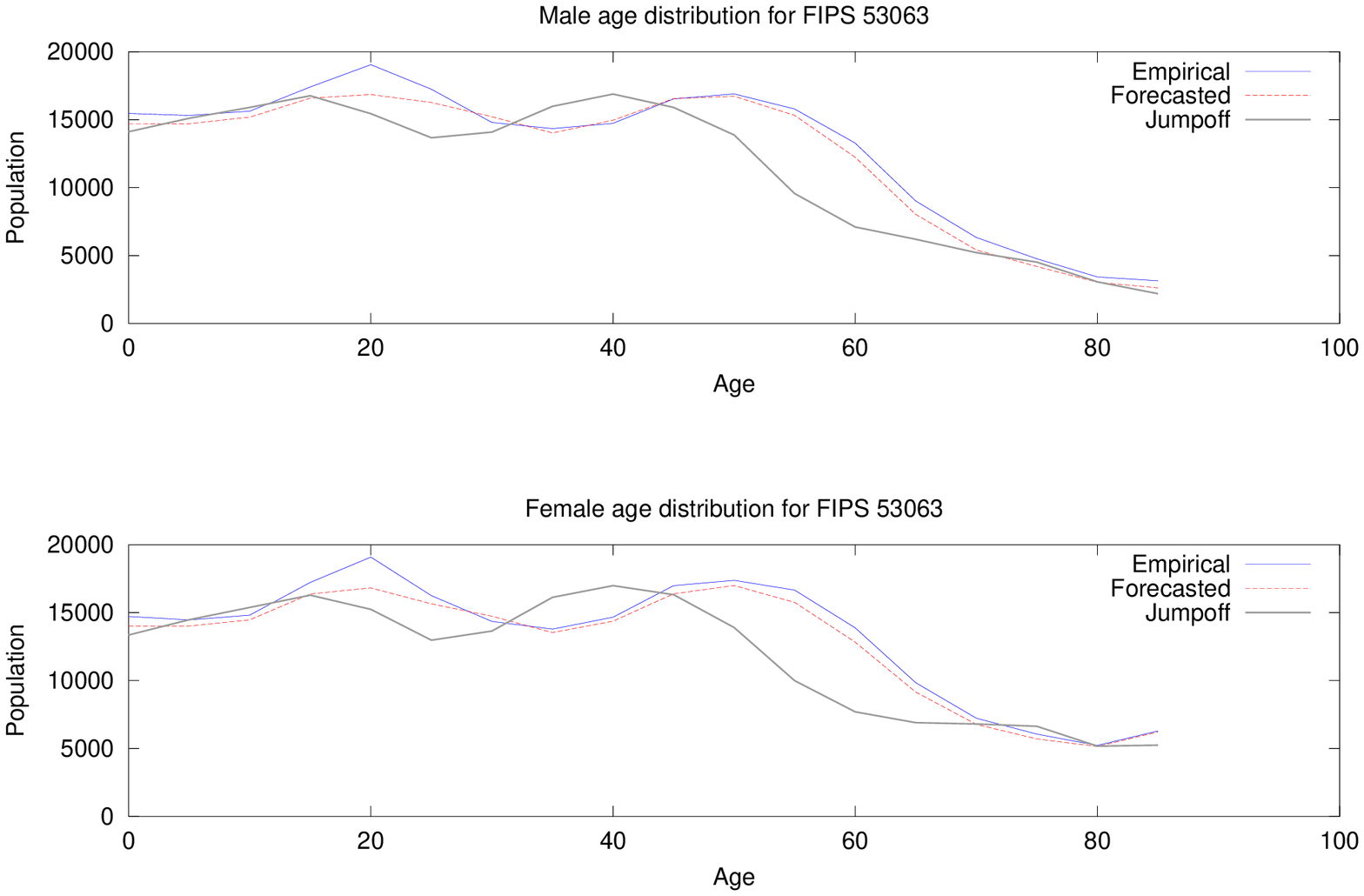}\caption{\label{fig:Spokane-County}Spokane County}
\end{figure}

\par\end{flushleft}

\newpage{}
\newpage{}
\newpage{}
\newpage{}
\newpage{}

\pagebreak{}

\end{document}